\documentclass[10pt]{article}

\usepackage[utf8]{inputenc}
\usepackage[T1]{fontenc}
\usepackage{epsfig}
\usepackage{subfigure}
\usepackage{graphicx}
\usepackage{algorithm}
\usepackage{algorithmic}
\usepackage{mathrsfs}
\usepackage{amsfonts}
\usepackage{float}
\usepackage{amssymb,amsmath}
\usepackage{indentfirst}
\usepackage{bm}
\usepackage{dsfont}
\usepackage{makecell}
\usepackage{theorem}
\usepackage{multirow}
\usepackage{array}
\usepackage{multirow}
\usepackage{subfigure}
\usepackage{longtable}
\usepackage{threeparttable}
\usepackage[table]{xcolor}
\usepackage{booktabs}
\usepackage{todonotes}

\newif\iftwoc
\twocfalse

\newif\ifpfig
\pfigtrue

\iftwoc

\makeatletter
\def\@normalsize{\@setsize\normalsize{12pt}\xpt\@xpt
\abovedisplayskip 12pt plus3pt minus7pt\belowdisplayskip \abovedisplayskip
\abovedisplayshortskip \z@ plus3pt\belowdisplayshortskip 6.5pt plus3.5pt
minus3pt\let\@listi\@listI}

\def\subsize{\@setsize\subsize{12pt}\xipt\@xipt}

\def\section{\@startsection {section}{1}{\z@}{24pt plus 2pt minus 2pt}
{12pt plus 2pt minus 2pt}{\large\bf}}

\def\subsection{\@startsection {subsection}{2}{\z@}{12pt plus 2pt minus 2pt}
{12pt plus 2pt minus 2pt}{\subsize\bf}}
\makeatother

\else

\renewcommand{\baselinestretch}{1.2}

\fi

\newlength{\wone}
\newcommand{\eq}{\begin{equation}}
\newcommand{\en}{\end{equation}}

\def\DIC{{\hbox{\rm\kern.2em\raise.36ex%
\hbox{$\scriptstyle |$}\kern-.4em C}}}
\def\SIC{{\hbox{\scriptsize\rm\kern.2em\raise.4ex%
\hbox{$\scriptscriptstyle |$}\kern-.4em C}}}

\def\DIQ{{\hbox{\rm\kern.2em\raise.4ex%
\hbox{$\scriptstyle |$}\kern-.4em Q}}}
\def\SIQ{{\hbox{\scriptsize\rm\kern.2em\raise.4ex%
\hbox{$\scriptscriptstyle |$}\kern-.4em Q}}}

\def\DZZ{{\hbox{\sf Z\kern-.41em Z}}}
\def\SZZ{{\hbox{\scriptsize\sf Z\kern-.41em Z}}}

\begin{document}
\date{}
\title{\Large\bf PolySmart and VIREO @ TRECVid 2024 Ad-hoc Video Search}
\author{
	Jiaxin Wu$^\dagger$,  Chong-Wah Ngo$^\star$, Xiao-Yong Wei$^{\mathsection\dagger}$, Qing Li$^\dagger$ 
    \vspace{0.08in}
   \\ {\em $^\dagger$Department of Computing, The Hong Kong Polytechnic University}
     \\    {\em $^\star$School of Computing and Information Systems, Singapore Management University}
          \\    {\em $^\mathsection$Department of Computer Science, Sichuan University}
   \\ \{jiaxwu, cs007.wei, prof.li\}@polyu.edu.hk, 
   cwngo@smu.edu.sg\\
}
\maketitle 
\section*{\centering Abstract}
\indent In this paper, we summarize our run and results submitted for the Ad-hoc Video Search (AVS) task at TRECVid 2024 \cite{2023trecvidawad}.

\vspace{0.2cm}
\label{abs:avs}\noindent\textbf{Ad-hoc Video Search (AVS):}This year, we explore generation-augmented retrieval for the TRECVid AVS task. Specifically, the understanding of textual query is enhanced by three generations, including Text2Text, Text2Image, and Image2Text, to address the out-of-vocabulary problem. Using different combinations of them and the rank list retrieved by the original query, we submitted four automatic runs. For manual runs, we use a large language model (LLM) (i.e., GPT4) to rephrase test queries based on the concept bank of the search engine, and we manually check again to ensure all the concepts used in the rephrased queries are in the bank. The result shows that the fusion of the original and generated queries outperforms the original query on TV24 query sets. The generated queries retrieve different rank lists from the original query. We briefly summarize our runs as follows:
\begin{itemize}
  \item[\textbullet]\textit{F\_M\_C\_D\_PolySmartAndVIREO.24\_1}: This automatic run reaches the mean xinfAP$=0.294$ on the main task using caption-to-video retrieval. 
  \item[\textbullet]\textit{F\_M\_C\_D\_PolySmartAndVIREO.24\_2}:
  This automatic run attains the mean xinfAP$=0.283$ on the main task by image-to-video search.
  \item[\textbullet]\textit{F\_M\_C\_D\_PolySmartAndVIREO.24\_3}: This automatic run obtains the mean xinfAP$=0.277$ on the main task. It combines the rank lists of image-to-video and caption-to-video searches with equal weights.
  \item[\textbullet]\textit{F\_M\_C\_D\_PolySmartAndVIREO.24\_4}: This automatic run attains the mean xinfAP$=0.277$ on the main task. It ensembles the results of run \textit{F\_D\_C\_D\_PolySmartAndVIREO.24\_3}, BLIP2 \cite{Li2023BLIP2BL}, CLIP \cite{radford2021clip} and Imagebind \cite{girdhar2023imagebind}.
  \item[\textbullet]\textit{F\_M\_N\_D\_PolySmartAndVIREO.24\_1}: This novelty run is based on the embedding-based search of our interpretable embedding model (ITV) \cite{wu2023ITV} with generated captions. As a result, this run attains mean xinfAP$=0.0.216$ for the main task. 
  \item[\textbullet]\textit{M\_M\_C\_D\_PolySmartAndVIREO.24\_2}: This manual run is based on the same system with the same settings presented in the run \textit{F\_M\_C\_D\_VIREO.24\_2} with manual selected visual queries. This run obtains 0.274 in the main task.
  \item[\textbullet]\textit{M\_M\_C\_D\_PolySmartAndVIREO.24\_3}: This manual run is based on the same setting presented in the run \textit{F\_M\_C\_D\_VIREO.24\_3} but with manual queries. It decreases the automatic result from 0.277 to 0.280.
  
 
\end{itemize}

\section{Ad-hoc Video Search (AVS)}

\begin{figure*}[t]
    \centering
    \includegraphics[width=1\linewidth]{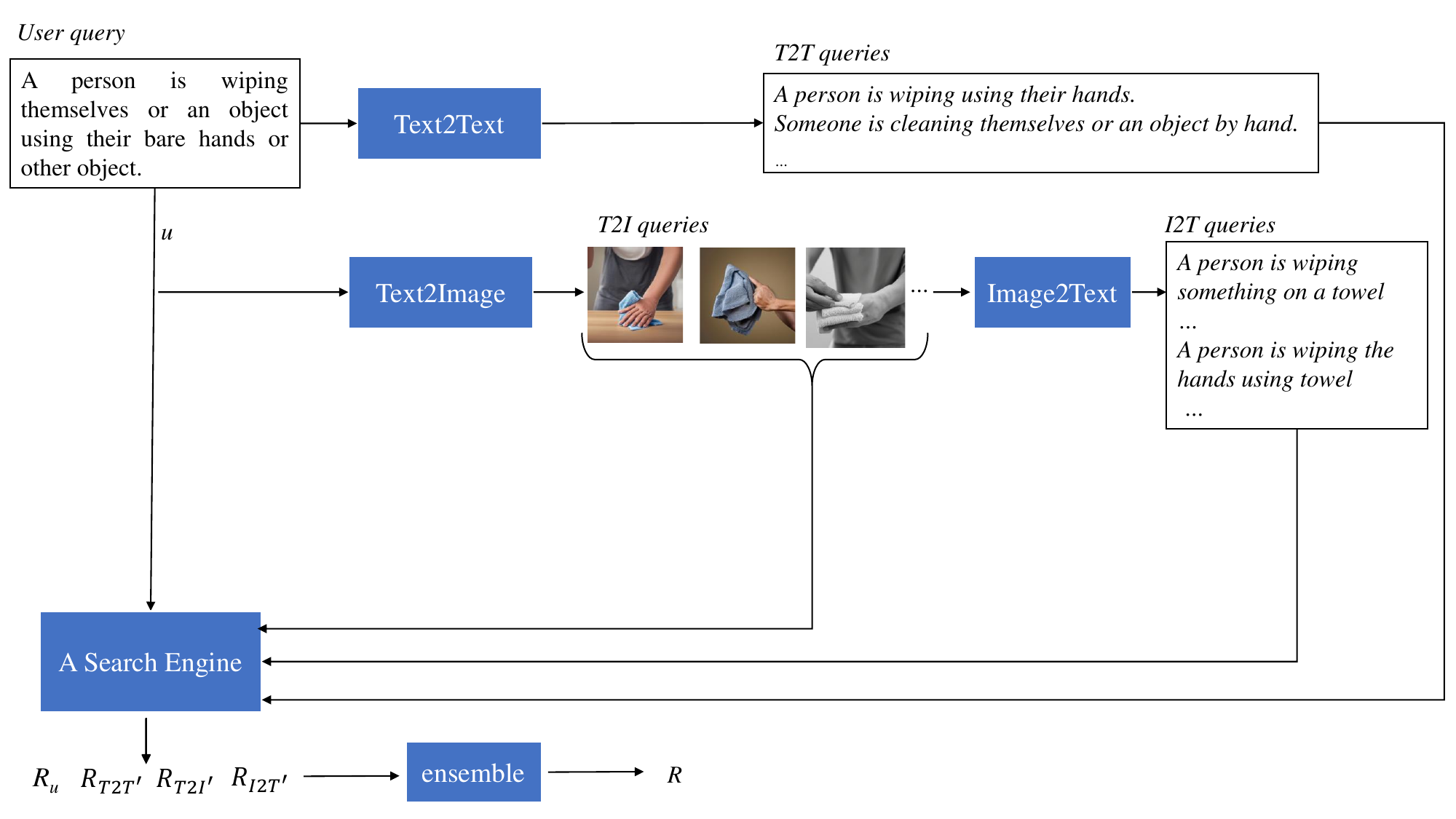}
    \caption{The overview of the generation-augmented retrieval plug-in module.}
    \label{fig:framework}
\end{figure*}

Ad-hoc video search \cite{2023trecvidawad} allows users to input ad-hoc queries to the search system to retrieve related videos. Due to the open-vocabulary setting, the existing AVS methods \cite{vireo2023,vireo2022,vireo2021,chongwahngo2010trecvid,chongwahngo2008trecvid,chongwahngo2005trecvid} usually fail badly on out-of-vocabulary (OOV) queries, logical queries and queries with spatial constraints. In light of the recent advance of the generative models on understanding natural language and transform it to images, we propose a retrieval-augmented query understanding plug-in module on top of our existing AVS model for TRECVid AVS task this year. The overview of the module is shown in Figure \ref{fig:framework}. We perform three kinds of transformations to address the OOV problem, including Text2Text (T2T), Text2Image (T2I), and Image2Text (I2T) transformations. Through these three generations, we replace the OOV words with existing concepts. Furthermore, we explicitly translate textual OOV words to visual concepts to narrow the cross-modality challenge for the retrieval system.

\section{Understanding the Query by generations}
Given an AVS query, it is input to three cross-model generative models (i.e., T2T, T2I, and I2T) as a prompt to generate texts/images/captions to rephrase the query to different modalities and with different context.  For the T2T transformation, an LLM (e.g., LlaMA 3 \cite{li2025llama}) accepts the existing knowledge of the search system (e.g., concept bank of a concept-based model) as context, and rephrases the OOV words with synonyms in the concept bank. For example, for the query \textit{Find shots of people standing in line outdoors}, "standing in line" is a OOV phrase, Through T2T generation, the query is changed to \textit{Find shots of people lineup outdoors}. For the T2I transformation, the textual query is input to a text-to-image generation model (e.g., stable diffusion \cite{rombach2021stablediffusion}) to map the textual words to visual concepts. By performing text-to-image generation, the OOV words are eliminated, and the logical and spatial constrains in the query are encoded in the images,  For the I2T transformation, we use an image captioning model (e.g., BLIP-2 \cite{Li2023BLIP2BL}) to further transform the generated images back to text with rich context. After three generations, the textual query along with three types of generated queries are input our existing AVS model \cite{improvedITV} to obtain four rank lists. We fuse these four rank lists with linear combination using equal weights.

\section{Results analysis}

\begin{figure*}[t]
    \centering
    \includegraphics[width=1\linewidth]{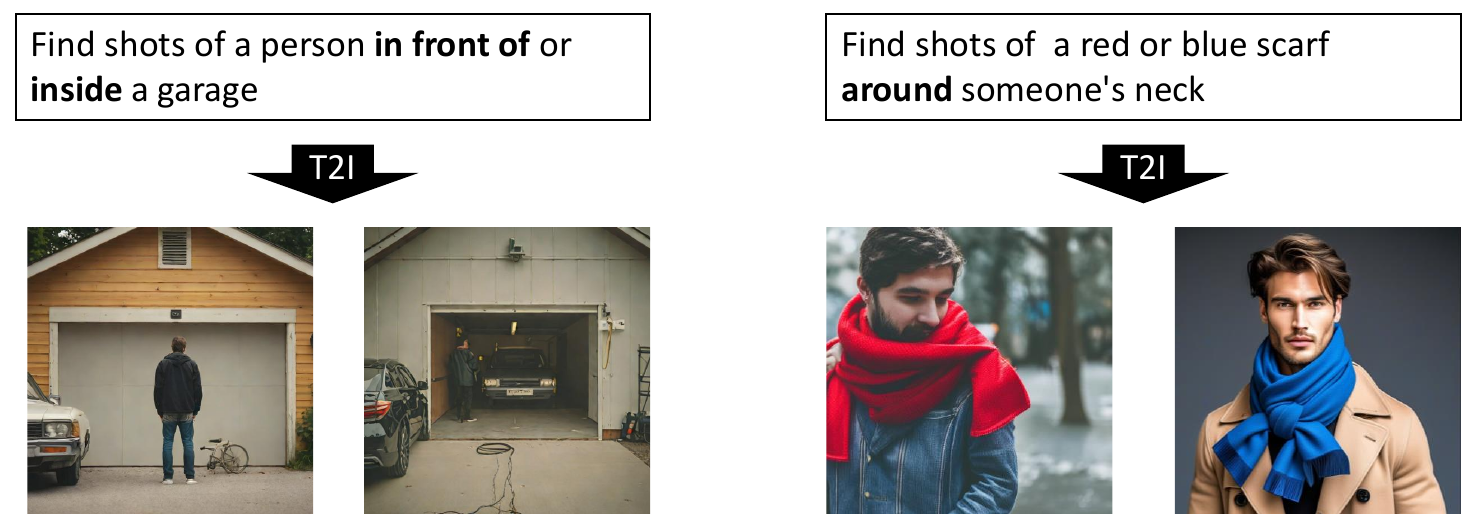}
    \caption{Example of an improved query by the generation.}
    \label{fig:T2I_examples}
\end{figure*}

Figure \ref{fig:T2I_examples} shows two examples of T2I generations of queries with spatial and logical constraints. For the query having restricted spatial constraints of a person and a garage, the T2I queries could represent the search intension accurately. For the query looking for blue or red scarf, a scenario of a man wearing blue scarf and a image of a man with red scarf around the neck are generated. 

\begin{table}[h]
\caption{Retrieval performance comparison on TRECVid AVS main query set (tv24).}
\label{tab:tv24_performances}
\centering
\begin{tabular}{cccccccc}
\bottomrule
\textbf{Run ID} & \textbf{Query} & \textbf{xinfAP}  \\
\bottomrule
Run 1 & original + T2T + T2I + I2T queries & \textbf{0.294} \\
Run 2 & T2T query & 0.283 \\
Run 3 &  I2T query & 0.277 \\
Run 4 &  original query & 0.277 \\
Run Novelty & T2I query &0.216 \\
\bottomrule
\end{tabular}
\end{table}

In this year's AVS benchmarking, the evaluation is conducted on the V3C2 dataset \cite{v3c2} with 30 queries. The overall results are shown in Table \ref{tab:tv24_performances}. The fusion of the original query and three types of generated queries obtains the best result. The I2T query obtains a similar performance to the original query, and the T2T outperforms them by 0.05. The T2I queries get the lowest performance.

\begin{figure*}[t]
    \centering
    \includegraphics[width=1\linewidth]{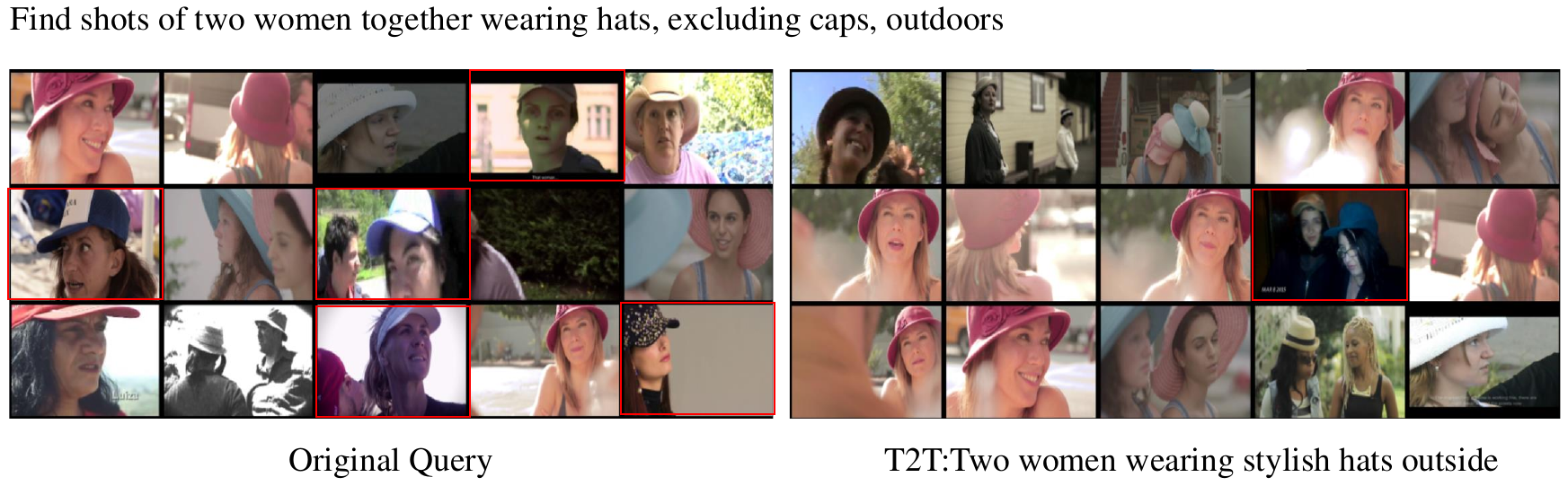}
    \caption{Example of an improved query by the generation.}
    \label{fig:improvedQuery}
\end{figure*}

\begin{figure*}[t]
    \centering
    \includegraphics[width=1\linewidth]{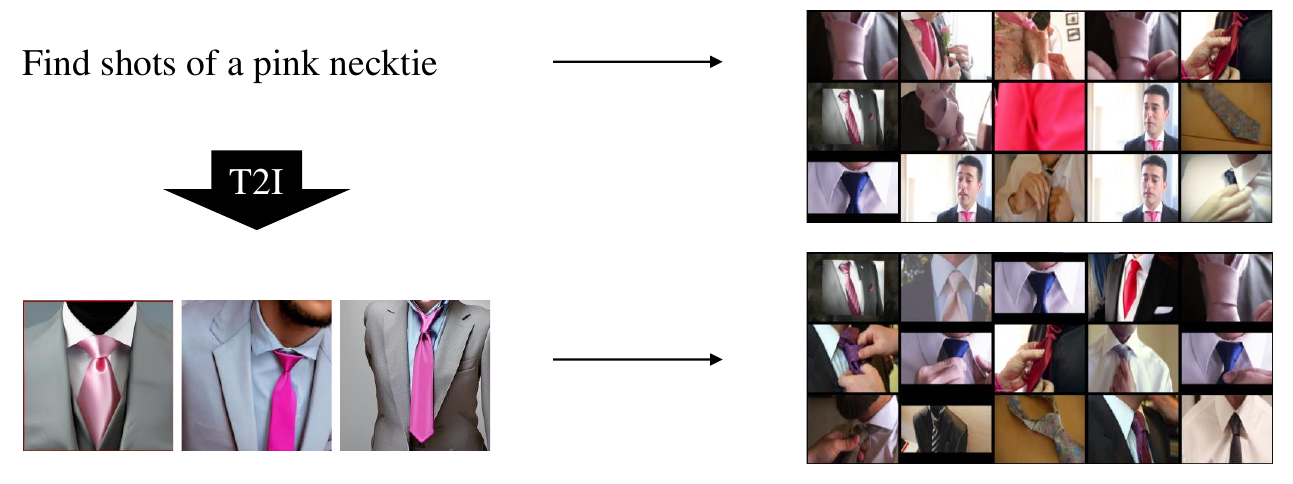}
    \caption{Example of an improved query by the generation.}
    \label{fig:weakImage}
\end{figure*}

Figures \ref{fig:improvedQuery} and \ref{fig:weakImage} visualizes a performance-improved query and a performance-dropped query, respectively. Figures \ref{fig:improvedQuery} compares the top-15 videos found by the original query (i.e., \textit{Find shots of two women together wearing hats, excluding caps, outdoors}) and the T2T query  (i.e., \textit{Find shots of two women wearing stylish hats outside}). As can be seen, the videos retrieved from the original query are mixed with people wearing hats or caps. With the T2T transformation, the main proportion of top-15 videos become stylish hats. On the other hand, Figure \ref{fig:weakImage} shows a performance-dropped query due to bad image-to-video retrieval on colour. As can be seen, the generated images are good and trustworthy compared to the original query. However, the retrieved videos have the semantic meaning with the generated images (i.e.,\textit{ a person wearing a suit and a tie}) but not pink necktie. Thus, the performance is dropped seriously. 

\begin{figure*}[t]
    \centering
    \includegraphics[width=1\linewidth]{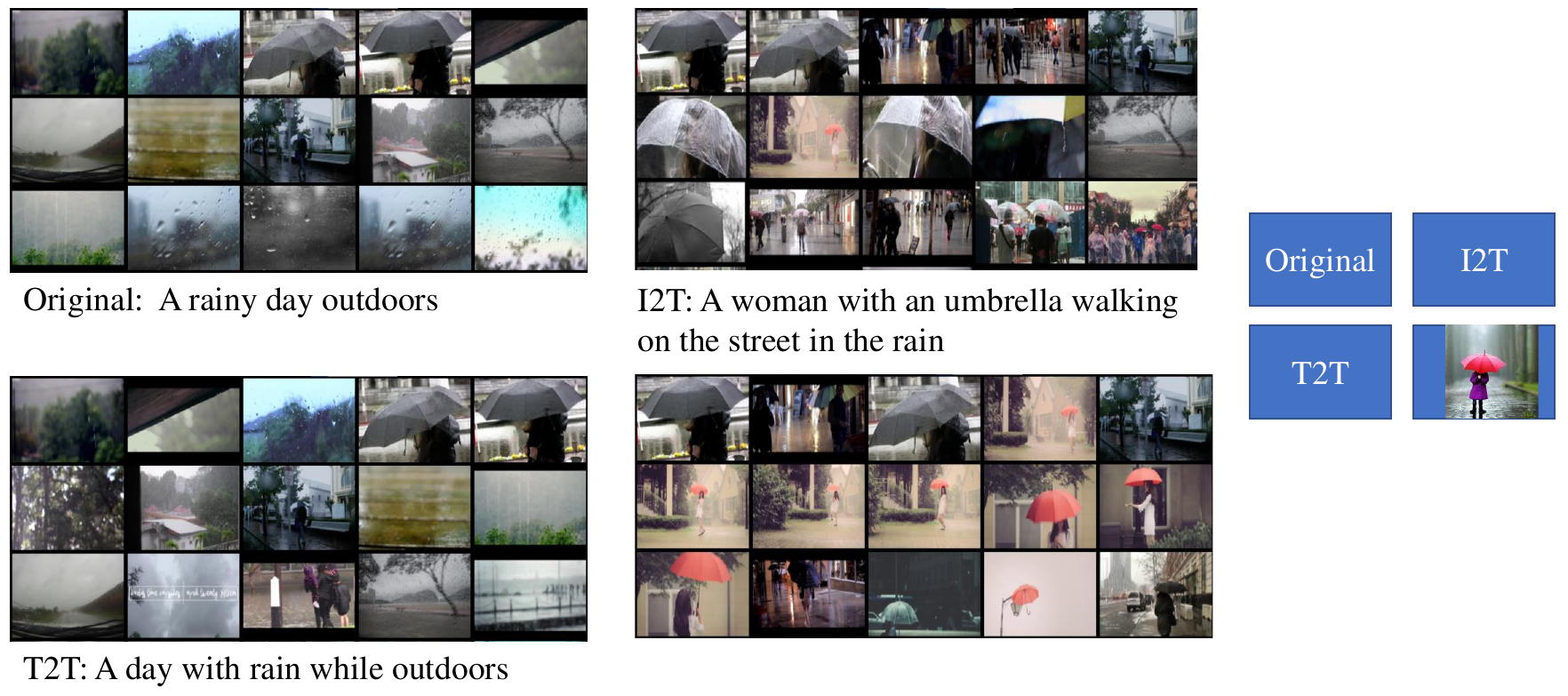}
    \caption{Example of an improved query by the generation.}
    \label{fig:resultComparison}
\end{figure*}

We also found that the results retrieved by the T2I queries have the biggest difference from the original query. An example is shown in Figure \ref{fig:resultComparison} for the query \textit{ Find shots of a rainy day outdoors}. The original and the T2T queries tend to retrieve videos without a person and an umbrella. In contrast, the T2I and the I2T queries mainly find videos of a person holding an umbrella and walking in the rain, which resembles the content represented by the T2I queries.

The manual queries are shown in Table \ref{tab:manual_query}. We manually pick one from the generated query as the manual query. Manually picking visual queries do not improve performance, while manual I2T queries boost automatic queries by 0.03. 

\begin{table}[]
\caption{manually generated queries for tv24}
\label{tab:manual_query}
\resizebox{\linewidth}{!}{

\begin{tabular}{l|l|l|l|l}
    & \textbf{Original query  }                                                                   & \textbf{Manually formulated query}                                           & \textbf{Manually selected T2T queries  }                           & \textbf{Manually selected I2T queries   }                                      \\
    \toprule
751 & A bald man with glasses                                                            & A bald man wearing glasses                                          & 751 A bald man wearing glasses                            & 751 a man with glasses and a bald head                                \\
752 & A rainy day outdoors                                                               & A rainy day outdoors in street                                      & 752 A day with rain while outdoors                        & 752 a woman with an umbrella walking on the street in the rain        \\
753 & A pink necktie                                                                     & A pink colored necktie                                              & 753 A pink-colored necktie                                & 753 a man in a suit and a pink tie`                                   \\
754 & A white sweater                                                                    & A white long sleeves sweater made of wool, cotton, synthetic fibers & 754 A white knitted sweater                               & 754 a woman standing on a street wearing a white sweater              \\
755 & A person is wiping themselves or an object using their bare hands or other object. & A person is wiping something with a towel                           & 755 A person is wiping using their hands                  & 755 a person is wiping something on a towel                           \\
756 & A man is putting on a jacket or a t-shirt                                          & A man is putting on a jacket                                        & 756 A man is putting on a jacket or a t-shirt             & 756 a man putting his jacket on while standing                        \\
757 & A man wearing a checked shirt                                                      & A man wearing a checked-pattern shirt                               & 757 A man with a checked pattern shirt                    & 757 a man in a checked shirt is looking at the camera                 \\
758 & A woman wearing a floral top or dress                                              & A woman wearing a floral dress                                      & 758 A woman wearing a dress or shirt with floral patterns & 758 a pretty woman wearing a floral dress                             \\
759 & People inside an airport terminal                                                  & People inside an airport terminal with Flight arriving or departure & boards                                                    & 759 people walking down an airport terminal with luggage              \\
760 & A man inside a workshop                                                            & A man inside a workshop holding tools                               & 760 A man crafting something inside a workshop            & 760 a man standing in a workshop with tools                           \\
761 & A traffic light seen at an intersection of a road or street                        & A traffic light seen at an intersection                             & 761 A traffic light seen at a road crossing               & 761 traffic lights at a street intersection with cars going both ways \\
762 & A map seen on a wall indoors                                                       & A map hanging on a wall indoors                                     & 762 A map hanging on a wall inside                        & 762 the world map is on a wall                                        \\
763 & At least two persons in a hallway are seen walking                                 & two persons in a hallway are seen walking indoors                   & 763 A group of people seen walking in a corridor          & 763 two people walking down the hallway in a building                 \\
764 & An adult is sitting in a glass walled building                                     & An adult is sitting in a building with glass walls                  & 764 An adult sitting in a glass-walled building           & 764 a man is sitting on a chair inside of a glass building            \\
765 & An adult is wrapped in a blanket                                                   & An woman or man is wrapped in a blanket                             & 765 An adult wrapped in a blanket                         & 765 a person that is wrapped in a blanket                             \\
766 & A person holding a pen                                                             & A person writing                                                    & 766 An individual holding a writing pen                   & 766 a person is writing on something with a pen                       \\
767 & A seated person reading from a paper or book outdoors during daytime               & A person sitting and reading a book outdoors during daytime         & 767 A person sitting and reading a book outside           & 767 a man sits outside while reading a book                           \\
768 & A woman wearing a silver necklace around her neck                                  & A woman wearing a silver necklace around her neck                   & 768 A woman wearing a silver necklace around her neck     & 768 a woman wearing multiple silver chains around her neck`           \\
769 & Two or more persons indoors with coffee cups or mugs seen with them.               & Two persons inside a coffee shop with cups or mugs.                 & 769 Two or more individuals holding coffee cups inside    & 769 three persons sitting at a table with cups of coffee`             \\
770 & Two women together wearing hats, excluding caps, outdoors                          & Two women wearing hats outdoors                                     & 770 Two women wearing stylish hats outside                & 770 two beautiful ladies are wearing hats outdoors`    \\              \bottomrule
\end{tabular}
}
\end{table}
\section{Conclusion}
Our study this year aims to explore generation-augmented queries to enhance the model's understanding of AVS queries. We find that the fusion of the original and generated queries could obtain better performance than the original query, and the generated queries could address the OOV words and logical/spatial queries. However, the current image-to-video retrieval should be improved with query attention. The manual query does not bring significant improvement to the original query. 

\section{Acknowledgments}
This research project is supported by the Ministry of Education, Singapore, under its Academic Research Fund Tier 2 (Proposal ID: T2EP20222-0047) and the National Natural Science Foundation of China (Grant No.: 62372314). Any opinions, findings and conclusions or recommendations expressed in this material are those of the author(s) and does not reflect the views of the Ministry of Education, Singapore, nor the National Natural Science Foundation of China.

\bibliographystyle{IEEEtran}
\bibliography{trec}
\end{document}